\documentclass[11pt,titlepage,a4paper]{article}
\usepackage{bozhomac}	
\usepackage{bozhlogo}	
\usepackage{cite}	

\title{\bfseries	\vspace*{-1.678902345in}
{\huge   Deviation equations\\[1ex]
	 in spaces with torsion}
}

\vspace{1.7ex}

\author{
Bozhidar Z.\ Iliev
\thanks{Laboratory of Mathematical Modeling in Physics,
Institute for Nuclear Research and \mbox{Nuclear} Energy,
Bulgarian Academy of Sciences,
Boul.\ Tzarigradsko chauss\'ee~72, 1784 Sofia, Bulgaria}
\thanks{E-mail address: bozho@inrne.bas.bg}
\thanks{URL: http://theo.inrne.bas.bg/$\sim$bozho/}
\\
\fbox{Sawa S.\ Manoff}
\thanks{Laboratory of Elementary particles,
Institute for Nuclear Research and \mbox{Nuclear} Energy,
Bulgarian Academy of Sciences,
Boul.\ Tzarigradsko chauss\'ee~72, 1784 Sofia, Bulgaria}
\thanks{Prof.\ Sawa Manoff passed away on May 27, 2005}
\thanks{URL: http://theo.inrne.bas.bg/elpart/SavaManov/smanoff.html}
}

\date{
 \vspace{2.27ex}\ShortTitle{Deviation equations in spaces with torsion}
	\\[0.27ex]
 \vspace{3.27ex}
\small
	\begin{tabular}{r@{$\colon\to~$}l}
 \vspace{0.09ex} Basic ideas	& 1980--1981	\\[0.09ex]
 \vspace{0.09ex} Last update	& June 28, 2005	\\[0.09ex]
 \vspace{0.27ex} Produced	& \fbox{\today}	\\[0.27ex]
	\end{tabular} \\[1.27ex]
\normalsize
	\begin{tabular}{r@{$\colon~$}l}
\vspace{0.27ex}
\normalsize\sffamily\bfseries
 http://www.arXiv.org e-Print archive No. &
\normalsize\sffamily\bfseries 	gr-qc/0507002	\\[1.27ex]
	\end{tabular} \\[-0.27ex]
\bfseries Published in\\\bfseries\itshape
Proceedings of the 5-th Soviet (USSR) Gra\-vity Conference\\
``Modern  theoretical and experimental problems of relativity theory\\
and gravitation'', Moscow Univ.\ Press, Moscow, 1981, p.~122 (In Russian)\\
\normalfont
 \vspace{4.27ex}{\Huge\BOZHO}	\\[4.27ex]
 \vspace{0.27ex}\Subject{General relativity, Differential geometry} \\[2.27ex]
	\begin{tabular}{r@{\hspace{0.512em}}|@{\hspace{0.512em}}l}
 \vspace{0.27ex}	
&
 \vspace{0.27ex} 
	\end{tabular} \\[1.27ex]
\vspace{0.27ex}
 							\\[0.27ex]
}

\begin{document}		

\renewcommand{\thepage}{\roman{page}}

\renewcommand{\thefootnote}{\fnsymbol{footnote}} 
\maketitle				
\renewcommand{\thefootnote}{\arabic{footnote}}   




\renewcommand{\thepage}{\arabic{page}}



\renewcommand{\theequation}{\arabic{equation}}

\vspace*{3ex}
	\begin{center}
	\begin{minipage}{0.75\textwidth}
\textbf{Abstract.}
The most general form of the deviation equations in spaces with linear
connection with arbitrary torsion is derived.
 \\[5ex]
	\end{minipage}
	\end{center}

{\Large\bfseries 1.}
	The deviation equations of two trajectories (paths) in a Riemannian
space $V_n$ find applications in number of problems in the gravity
theory~\cite{Yano/LieDerivatives}. The method of their derivation can be
generalized to spaces $L_n$ with affine (linear) connection $\nabla$, which
is, of course, applicable to spaces with torsion $U_n$.

{\Large\bfseries 2.}
	For any vector $\xi\in T_x(L_n)$ in the space $T_x(L_n)$ tangent to
$L_n$ at $x\in L_n$, the following identity holds:
	\begin{equation}	\label{1}
\frac{D^2 \xi}{\od s}
=
\Hat{R}(u,\xi) + C_{1}^{1} (F\otimes D\xi) +
\frac{D \Hat{T}(u,\xi))}{\od s} -
T(F,\xi)+
\mathfrak{L}_\xi V -
\frac{D \mathfrak{L}_\xi u}{\od s} -
C_{1}^{2} (Du\otimes \mathfrak{L}_\xi u) .
	\end{equation}
Here:
$ \frac{D}{\od s} := u^i\nabla_{E_i}$
 $u^i:=\frac{\od x^i(s)}{\od s}$, $\nabla_{E_i}$ is the covariant derivative
along $E_i$ with $\{E_i$\} being a frame in a neighborhood of $x$,
$u=u^iE_i$,
 $Du:=(\nabla_{E_i}u)\otimes \Theta^i$ with $\{\Theta^i\}$ being a frame dual
to $\{E_i\}$ ($\Theta^i(E_j)=\delta^i_j$),
$s$ is a parameter of a path $\gamma(s)\in L_n$, $s\in\field[R]$,
 $\Hat{R}$ is the curvature operator with
\(
\Hat{R}(u,\xi) v
:=
\bigl( \nabla_u\nabla_\xi - \nabla_\xi\nabla_u - \nabla_{[u,\xi]} \bigr) v
=R_{jkl}^{i}v^ju^k\xi^l E_i,
\)
$ F:=\frac{D u}{\od s} = u^i\nabla_{E_i}u$,
$\Hat{T}$ is the torsion operator with
\(
\Hat{T}(u,\xi):=\nabla_u\xi-\nabla_\xi u - [u,\xi]
=
T_{jk}^{i} u^j\xi^k E_i ,
\)
 $\mathfrak{L}_\xi$ is the Lie derivative along $\xi$,
 and $[u,\xi]:=u\xi - \xi u$.
The identity~\eref{1} in component form reads
	\begin{equation}	\label{2}
\frac{\bar{D}^2 \xi^k}{\od s}
=
R_{ijl}^{k} u^iu^j \xi^l + \xi_{|j}^{k}F^j +
u^j \frac{\bar{D} (T_{jl}^{k} \xi^l) }{\od s}
+\mathfrak{L}_\xi F^k -
T(F,\xi)+
\frac{\bar{D} (\mathfrak{L}_\xi u)^k}{\od s} -
u_{|i}^{k} \mathfrak{L}_\xi u^i ,
	\end{equation}
where
$ \frac{D v}{\od s} := u^i\bar{\nabla}_{E_i} $,
 $\nabla_{E_i}v =: (\bar{\nabla}_{E_i}v^j) E_j$
and $v^i_{|j}:=\nabla)_{E_j}v^i$ is the covariant derivative of the
components $v^i$ along $E_i$.

{\Large\bfseries 3.}
	If we impose some additional restrictions, which in certain cases may
be first integrals of the deviation equation, on the quantities entering
into~\eref{1} (e.g.\ $F=0$, $\mathfrak{L}_\xi u=0$, $\mathfrak{L}_\xi F=-F$),
then we get corresponding deviation equations in space with torsion $U_n$, in
which the vector $\xi$ is considered as an infinitesimal one.

\addcontentsline{toc}{section}{References}
\bibliography{bozhopub,bozhoref}
\bibliographystyle{unsrt}
\addcontentsline{toc}{subsubsection}{This article ends at page}

\end{document}